\documentclass[aps,prl,showpacs,twocolumn,groupedaddress]{revtex4}

\usepackage{graphicx}

\begin{document}

\title{Finding the structure of phosphorus in the phase IV}



\author{Takahiro Ishikawa}%
 \email{ishikawa@aquarius.mp.es.osaka-u.ac.jp}
\author{Hitose Nagara}%
 \email{nagara@mp.es.osaka-u.ac.jp}
\author{Koichi Kusakabe}%
 \email{kabe@mp.es.osaka-u.ac.jp}
\author{Naoshi Suzuki}%
 \email{suzuki@mp.es.osaka-u.ac.jp}

\affiliation{%
Division of Frontier Materials Science, Graduate School of Engineering Science, 
Osaka University, Toyonaka, Osaka 560-8531, Japan
}%

\date{\today}

\begin{abstract}
We have explored the unknown structure of the phosphorus in the phase IV 
(P-IV phase) 
based on the first-principles calculations using the metadynamics simulation 
method. 
Starting from the simple cubic structure, we found a new modulated structure 
of a monoclinic lattice. 
The modulation is crucial to the stability 
of the structure. 
Refining further the structure by changing the modulation period, 
we have found the structure which shows the X-ray powder 
pattern in the best agreement with the experimental one.
We can not exclude the possibility that the unknown structure 
of the phase IV of phosphorus is an incommensurately modulated one.
\end{abstract}


\maketitle


Recent progress in high-pressure physics strengthened 
our recognition on the variety of structures of materials. 
Unexpectedly interesting structures were found in the high-pressure 
experiments. The modulated structure is one of curious structures 
often found in the high-pressure phases of elements. 
Improvement of the high-pressure techniques has also made it possible to
 specify the structure stabilized only in a narrow pressure range. 
The modulation 
of a crystal was found in the group Vb elements including As, Sb 
and Bi~\cite{Mcm00,Shw03} 
and in group VIb elements including S~\cite{Hej05}, Se~\cite{Mcm04} 
and Te~\cite{Hej03}. 
Modulated structures were also reported in halogens, I~\cite{Tak03} 
and Br~\cite{Kum05}. 

Non availability of sufficient experimental data under very high pressure
constrains high pressure study. There, people often encounter difficulties 
in determination of the 
lattice structure only from the experimental data. 
The theoretical approach gives alternative information on the same 
problem, where accuracy in the determination of a crystal structure 
is known to be enough, if one utilizes the first-principles calculation. 
However, limitation in computational resources 
sometimes prohibits us to perform full search of the true structure. 

The phase IV of the phosphorus (P-IV) is one of the examples 
in which the structure has not been determined. 
Observation of the phase was first reported by Akahama {\it et al.}~\cite{Aka99} 
in 1999. They showed an appearance of 
the simple hexagonal (sh) phase, 
{\it i.e.} the phase V (P-V), which is stabilized above 137GPa. 
The third phase (P-III) appearing in the sequence of pressure-induced 
transformations at low temperatures is 
the simple cubic (sc) phase. 
As an intermediate phase between 
sc and sh, the P-IV was detected in the X-ray diffraction data.
Experimentally, however, the structure has not been identified. 
The ordinary Rietveld analysis starting from the knowledge only
of the monoclinic 
symmetry has not been successful owing to a probable complexity of the lattice.
The bcc structure (phase VI) is theoretically predicted~\cite{Ham00} 
and identified at even higher pressure 262GPa~\cite{Aka00}.
We need a guess of the lattice structure or a pseudo crystal. 

Several structures have been tested as candidates for the P-IV. 
Ahuja considered a structure of space group Imma~\cite{Ahu03}.
Ehlers and Christensen studied relative stability of Ba-IV structure 
of P, which is a kind of modulated structure, in the pressure range 
from 100 to 200 GPa~\cite{Ehl04}. 
In spite of these intensive studies, the structure of the P-IV 
has not been identified. 

To explore the P-IV, we adopted the following strategy in 
our theoretical study and used the metadynamics simulation in the first-principles 
calculations.  This trial was done with a relatively small simulation 
cell to reduce the computational time. The simulation was planned, 
however, to be able to detect possible signal of the structural 
phase transformation. 
For the obtained structure, we checked the relative stability against 
the sc and sh phases. 
Next we considered some model structures to find more refined structure. 
The structural optimization was done for each model structures. 
The calculated X-ray powder patterns of the 
optimum structure are compared with that of the experimental one.

In the study of the metadynamics simulation, which was first introduced by
Laio and Parrinello~\cite{Lai02,Mar03},
we use the Gibbs free energy (GFE) depending upon a shape of 
a simulation cell.
Following the prescription by Marto\v{n}\'ak~\cite{Mar03}, we 
consider the GFE, $G(\mathbf{h}^{t}) = G_o(\mathbf{h}^{t}) + G_g(\mathbf{h}^{t})$, 
where $G_g(\mathbf{h}^{t})$ is the artificial potential defined 
by the following 
Gaussian-type function,
\begin{equation}
{G_g}(\mathbf{h}^{t})
= \sum_{t'<t}\prod_{i,j}W\exp (-[\mathbf{h}^{t} - \mathbf{h}^{t'}]_{ij}^{2}/2{\delta h}^{2}).
\end{equation}
The superscripts $t$ and $t'$ denote the current meta-step and the previous 
one, respectively. 
The quantities $W$ and $\delta h$ represent the weight and the width of 
the Gaussian-type function, respectively. 
The matrix $\mathbf{h}$ is defined by the vectors defining 
the simulation cell and $\mathbf{h}=(\vec{a},\,\vec{b},\,\vec{c})$, where  
$\vec{a}$, $\vec{b}$ and $\vec{c}$ are lattice vectors.  
To eliminate the free rotation of the system, only the  
symmetric part of the matrix $\mathbf{h}$ is updated, which reduces the number of 
the independent variables to 6.

The update of the matrix $\mathbf{h}$ is made by the steepest descent method using the 
driving force $\mathbf{F}$ and regarding $\delta h$
 as stepping parameter. The driving force $\mathbf{F}$ is obtained as the
sum of the original driving force $\mathbf{F}_{o}=-\partial G_o/
\partial \mathbf{h}$, and the Gaussian driving force $\mathbf{F}_{g} 
= -\partial G_{g}/\partial \mathbf{h}$. The force $\mathbf{F}_{o}$ can be expressed
by an internal pressure tensor $\mathbf{p}$, external pressure $P$, and 
the matrix $\mathbf{h}$~\cite{Mar03}.
One step of updating $\mathbf{h}$ is defined as one meta-step.

At each meta-step, in order to equilibrate the system 
and to estimate $\mathbf{p}$, 
the conventional molecular dynamics (MD) simulations have 
to be done with the shape 
of the simulation cell fixed. 
The internal pressure tensor and the atomic positions at each meta-step 
can be taken from the output of any constant-pressure MD codes of the 
first-principles calculation~\cite{BarXX}.

The above artificial potential means that if the current $\mathbf{h}$ has been 
visited time after time, which occurs when the system is fluctuating around 
the local minimum of the GFE surface, $G_g$ accumulates to a large value and 
the well is gradually filled with the artificial potential $G_{g}$. 

For the simulation of phosphorus,
we used the density functional theory in a local density approximation and 
a norm-conserving pseudopotential, where we employed the expression 
of Perdew and Zunger~\cite{Per81} for the exchange and correlation energy functional. 
We checked the pseudopotential comparing the calculated equation of states with the
experiments in other phases~\cite{Aka99,Aka00,Ham00}.
We started with the cubic simulation cell whose edge was 4.26~\AA \ and 
8 phosphorus atoms are set at the positions which make the sc lattice.  
We performed the k-space integration using 8 $\times$ 8 $\times$ 8 mesh 
points in the first Brillouin zone and set the energy cut-off of the plane wave basis at 40 Ry. 
We set the external pressure at 120 GPa in the conventional constant-pressure  
MD, since the P-IV is observed at this pressure. 
In order to equilibrate the system, 
we ran this MD simulation for 200 steps at each meta-step and we calculated the average internal pressure tensor 
from the latter half of 100 steps. 
In order to perform this metadynamics simulation, we used the cluster
machines.  

\begin{figure}
\includegraphics[width=8.5cm]{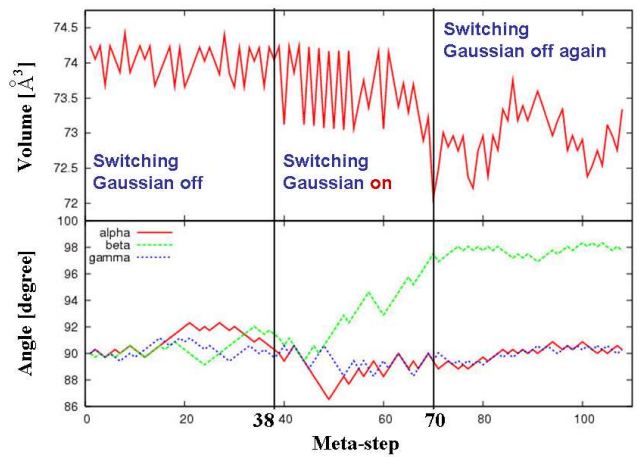}
\caption{\label{metasimulation} 
Evolution  of the simulation cell volume and the angles among lattice vectors. 
The Gaussian-type potential was switched on 
at 39th meta-step and off again after 71st meta-step.}
\end{figure}
\begin{figure}
\includegraphics[width=8.5cm]{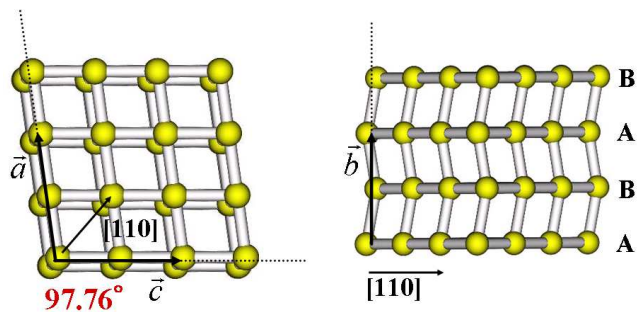}
\caption{\label{metastruct} 
A structure obtained by the first-principles metadynamics simulation. 
$\vec{a}$, $\vec{b}$ and $\vec{c}$ are lattice vectors of the simulation 
cell, and 
$a=4.22$~\AA, $b=4.15$~\AA, $c=4.22$~\AA, $\alpha = 90.86^{\circ}$, 
$\beta = 97.76^{\circ}$ and $\gamma = 90.26^{\circ}$.  When we look at the 
$ac$ plane along the $b$-axis, the planes are displaced alternatively in 
the direction [110].}
\end{figure}

Figure \ref{metasimulation} shows the evolution of the cell volume and the 
three angles among 
the lattice vectors of the simulation cell.  
First we ran the simulations with the Gaussian potential off, by setting 
$W=$ 0~mRy, for 38 meta-steps 
in order to check whether the sc structure is stable or not.   
During these initial meta-steps, the values of the angles and the volume were 
nearly maintained at those of the sc lattice. 
This means that the sc structure resides on a local minimum and it is 
separated from the other local minima by some barriers. 
We note that the metadynamics simulation with the Gaussian-type potential 
switched off is nearly equivalent 
to the conventional variable-cell MD simulations. 

In order to explore the metastable structures beyond the potential barrier
we switched  the Gaussian-type potential on at 39th meta-step with 
$W$ of 1~mRy and $\delta h$ of 20~m\AA. 
As a result, one of the three angles started to increase after around 
50th meta-step 
and the volume began to decrease dramatically. 
After those changes, we switched the Gaussian-type potential off  at 71st 
meta-steps again 
in order to check whether the system had already surmounted the barrier 
and had moved to a neighboring local minimum. 
If the system had not crossed the barrier yet, 
the angle and the volume would have returned to the starting 
values of the sc lattice, which are approximately 
$90^{\circ}$ and 74~\AA$^{3}$, respectively. 
After 71th meta-step, however, the volume fluctuated around 73~\AA$^{3}$ 
and three angles also continued to fluctuate around $90^{\circ}$, $98^{\circ}$ and 
$90^{\circ}$. 
This behavior shows that the sc structure transformed into another metastable 
one.

Figure \ref{metastruct} shows the structure obtained by the above run, where 
$\vec{a}$, $\vec{b}$ and $\vec{c}$ are the lattice vectors of the simulation cell. 
This structure has a unit cell with the lattice parameters $a=4.22$~\AA, $b=4.15$~\AA, $c=4.22$~\AA, $\alpha = 90.86^{\circ}$, 
$\beta = 97.76^{\circ}$ and $\gamma = 90.26^{\circ}$. 
The left hand side figure is the projection onto the $ac$ plane. 
It shows the distortion of the simulation cell from the cubic 
into the cell with an angle $97.76^{\circ}$.  When we look the lattice 
from a side,
we find a zigzag modulation of the $ac$ plane along the $b$-axis with displacement 
in the direction of [110]
as is shown in the ABAB.. in the right hand side figure.
This is an important feature. 
The ABAB$\cdots$ modulation pattern is crucial for the stability 
of the distortion of the angle $\beta$. 
When we removed the zigzag modulation pattern and 
performed the simulation for the relaxation of the structure, 
we observed that the structure returned to the initial sc structure.

\begin{figure}
\includegraphics[width=6.5cm]{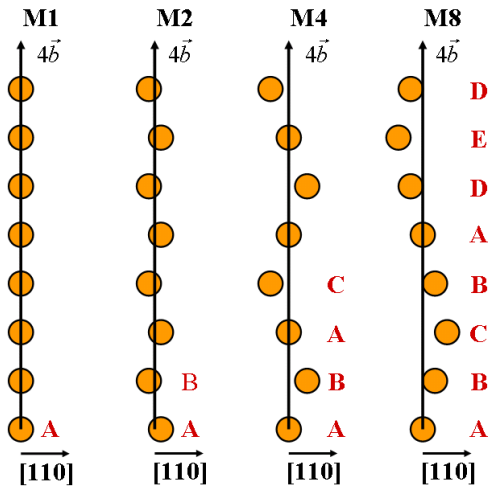}
\caption{\label{modulation} 
Modulated structures with different modulation periods along the $b$-axis. 
The pattern M1 is of a non-modulated structure, and M2 the structure obtained by our 
metadynamics simulation (ABAB$\cdots$). The modulation periods of 
M4 (ABAC$\cdots$) and M8 (ABCBADED$\cdots$) patterns are twice and 
four times as long as that of the M2, respectively.}
\end{figure}
\begin{table}
\caption{\label{totalenergy}
Comparison of the total energies per atom for a non-modulated structure (M1) 
and three modulated structures (M2, M4 and M8). 
We used the unit cell with the lattice vectors, $\vec{a}/2$, $4\vec{b}$ 
and $\vec{c}/2$ in the calculation of the total energy for all of these four structures.}
\begin{ruledtabular}
\begin{tabular}{cc}
      Structure                & Total energy [Ry/atom] \\ \hline
      M1                       & -13.1624 \\
      M2                       & -13.1638 \\
      M4                       & -13.1642 \\
      M8                       & -13.1637 \\
\end{tabular}
\end{ruledtabular}
\end{table}
\begin{figure}
\includegraphics[width=7.0cm]{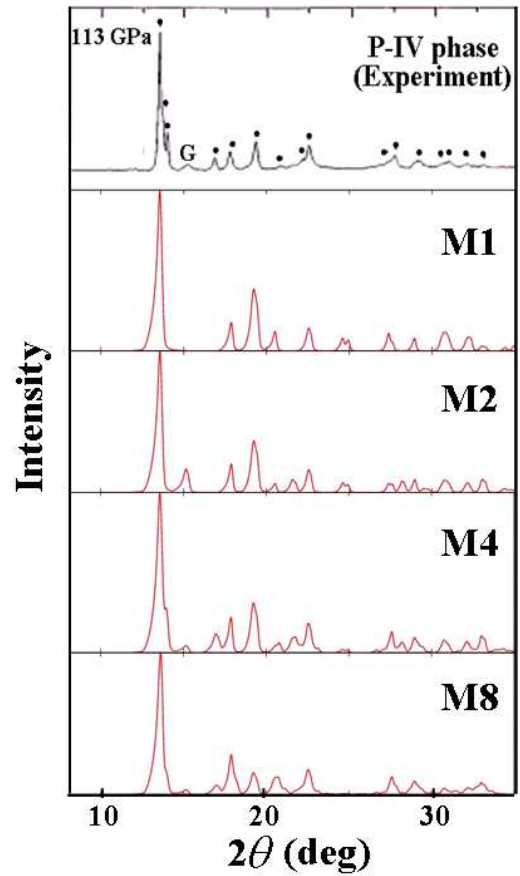}
\caption{\label{d-pattern2} 
Comparison of the experimentally obtained X-ray powder pattern 
of the P-IV~\cite{Aka99} 
with those of the M1, M2, M4 and M8 modulated structures. 
Theoretical powder patterns of our structures are obtained 
by the use of RIETAN-2000~\cite{Izu00}. 
In the experimental pattern of the P-IV, 
peaks are indicated by dots and G represents diffraction by 
a metal gasket. }
\end{figure}

The structure obtained by the metadynamics is a
modulation pattern with period consisting 
of two planes.
Our simulation, however, was performed using the system with 8-atoms in 
a simulation cell, and with the periodic boundary condition.  Hence there remains 
a question that the small simulation cell may limit the modulation period
to a shorter one.   In fact, the X-ray powder pattern of the zigzag 
modulated structure and that of the experimental one show some discrepancies.
To answer this question, we tried some more first-principles
calculations for the refinement of the structure.

We extended our study to two more structures 
which have commensurate modulations: an ABAC$\cdots$ 
and an ABCBADED$\cdots$. 
The modulation period of the ABAC$\cdots$ and that of the ABCBADED$\cdots$ 
 are twice and four times as long as 
that of the ABAB$\cdots$, respectively. 
These structures are denoted as M4 and M8 and shown in Fig. \ref{modulation}.  
The pattern  M1 is a non-modulated structure, and M2 is the structure obtained 
by the metadynamics simulation. 
We calculated the total energies and the X-ray powder patterns. 

To compare the total energy among these structures, 
we used the same unit cell with the lattice vectors, $\vec{a}/2$, $4\vec{b}$ 
and $\vec{c}/2$, where 
$\vec{a}$, $\vec{b}$ and $\vec{c}$ are those of the simulation 
cell obtained by the metadynamics simulation.
The 8 atoms were located only along the $b$ axis in the unit 
cell with displacements corresponding to the modulation pattern 
(Fig. \ref{modulation}). This choice of the unit cell avoids 
the numerical errors coming from the use of the
different size of the unit cells.
For the k-space integration, we used 16 $\times$ 4 $\times$ 16 mesh points 
in the first Brillouin zone. 
Amplitude of modulation for each structure was optimized by the relaxation 
of the atomic positions. 

Calculated total energies are listed in Table \ref{totalenergy}. 
All modulated structures, M2, M4, and M8 have lower energy 
per atom than the unmodulated structure, M1, which shows any modulation 
periods from M2 to M8 are more favorable than the M1. 
Among the above three modulated structures, 
the energy of the M4 is the lowest in our study
of the commensurate approximation.
Though the enthalpy of the M4 is very close to that of the sc, it is in fact 
lower than that of the sc at 120GPa, according to our results.
If we plot the total energy as a function of the modulation period and 
optimize the period as was done by Ehlers {\it et al.}~\cite{Ehl04}, 
an incommensurately modulated structure is expected. 

In figure \ref{d-pattern2} we compare the X-ray powder patterns of our 
structures with that of the experimental one. 
The experimental pattern of the P-IV (top figure) was obtained 
from Akahama {\it et al.}'s~\cite{Aka99}.
It shows the feature that the splits of the strongest peak 
at $2\theta = 13^{\circ}$ and the three peaks in the range from $2\theta = 17^{\circ}$ to $19^{\circ}$ are observed. 
In our M2 structure, these features are missing and unnecessary peaks exist
at $2\theta = 15^{\circ}$ and $25^{\circ}$. 
However, this disagreement is much improved with increase of the modulation
period. 
Intensities of the unnecessary peak at $2\theta = 15^{\circ}$ in the M2 structure,
 which was brought about by the zigzag modulation, and another unnecessary one at $2\theta = 
25^{\circ}$ in the same M2 structure, which appeared owing to the distortion 
of the simulation cell from the cubic, 
are decreased in the M4 
and M8 structures and the split of the strongest peak appears also in the M4 
and M8 structure. 
About the intensity of the three peaks in the range from $2\theta 
= 17^{\circ}$ to $19^{\circ}$, the X-ray powder pattern of the
M4 shows  most improved agreement with the experimental one among the 4 patters studied. 
From the comparison of the total energies and the X-ray powder patterns, 
we conclude that the modulation period is close to that of the M4 structure 
in the P-IV.

In this study, we explored the structure of phosphorus in the phase IV using 
the first-principles metadynamics simulation and 
identified the new structure.   
This structure is the monoclinic with the modulation pattern.
Furthermore we found the refined structure showing the best 
agreement with the X-ray powder pattern.
Although we have not fully studied the possibility of the 
incommensurate modulation, we conclude that phosphorus takes the modulated 
structure with possible incommensurate modulation.
The simple idea of 
Akahama {\it et al.}~\cite{Aka99}  
stating that the structure of the P-IV may be 
on the path from the sc to sh 
via monoclinic distortion along the [110] direction is partly supported 
because the
unmodulated structure is of space group $P2$ and of one atom per unit cell.
The structure of the P-IV is not so complicated as one suggested by and 
Ehlers {\it et al}~\cite{Ehl04}. 
The modulation stabilizes the monoclinic distortion of the lattice.
It is highly probable that the Vb, VIb, and VIIb group elements commonly show 
modulated structures in a narrow pressure range  when they undergo 
the pressure induced structural transition between simple stable structures.
\begin{acknowledgments}
The authors thank Dr. Nakamoto and Dr. Morimoto for valuable discussion on the
X-ray powder patterns.
Computations were done on the machines 
at the Institute for Molecular 
Science, Okazaki, Aichi, Japan. 
This work was partially supported 
by a Grant-in-Aid for Scientific  Research in Priority  Areas 
``Development of New Quantum Simulators and  Quantum Design'' 
(Nos.17064006 and 17064013)  and by a Computational Nanoscience program
``-Grid Application Research in Nanoscience-National Research Grid Initiative 
(NAREGI)'' of The Ministry of Education,  Culture,  Sports, Science, 
and Technology, Japan.
\end{acknowledgments}
\vspace*{-0.7cm}
\bibliography{paper2005prlx}

\end{document}